\documentclass{cimento}
\usepackage{amsmath,amssymb}
\usepackage[pdftex]{graphicx}
\usepackage{units}

\title{The end of the cosmic ray spectrum}

\author{Darko Veberi\v{c}\from{ins:ung},
for the Pierre Auger Collaboration\from{ins:pao}}

\instlist{
\inst{ins:ung}
Laboratory for Astroparticle Physics, University of Nova Gorica,
\\
Vipavska 13, SI-5000 Nova Gorica, Slovenia
\inst{ins:pao}
Observatorio Pierre Auger, Av.\ San Martin Norte 304, 5613 Malarg\"ue, 
Argentina
\\
{\normalfont Full author list: {\tt
http://www.auger.org/archive/authors\_2011\_05.html}}
\\
{\normalfont Email: {\tt auger\_spokespersons@fnal.gov}}
}

\PACSes{
\PACSit{96.50.S-}{Cosmic rays}
\PACSit{96.50.sd}{Extensive air showers}
\PACSit{96.50.Vg}{Energetic particles}
\PACSit{13.85.Tp}{Cosmic rays (high-energy interactions)}
}

\begin{document}

\maketitle

\begin{abstract}
Recent results from the Pierre Auger Observatory are presented, focusing 
on a measurement of the cosmic-ray energy spectrum above $10^{18}$~eV,
cosmic-ray composition, and the anisotropy in the cosmic ray arrival 
directions.

The flux of cosmic rays can be well described by a broken power-law, 
with a flattening of the spectrum above $4\!\times\!10^{18}$\,eV and a 
softening of the spectrum beginning at about $3\!\times\!10^{19}$\,eV.
The flux suppression at highest energies is consistent with the 
predictions of Greisen, and Zatsepin and Kuzmin. Longitudinal 
development of cosmic-ray air showers provides information on the mass 
of the primary particle. When compared to model predictions, our 
measurements of the mean and spread of the longitudinal position of the 
shower maximum are indicating a composition transition from light to 
heavier with increasing energy. For highest energies in our data-set we 
observe evidence for a correlation between the cosmic-ray arrival 
directions and the nearby extragalactic objects.
\end{abstract}

\section{Introduction}

The Pierre Auger Observatory~\cite{ea} was designed to measure 
properties of the extensive air showers produced by cosmic rays with 
ultra-high energies above $\unit[10^{18}]{eV}$. Since the occurrence of 
these rare events is of the order of magnitude of 1 per $\unit{km^2}$ 
per century, the Observatory has a large aperture in order to gather a 
statistically significant sample. The Observatory is featuring 
complementary detection techniques to lessen some of the systematic 
uncertainties associated with deducing properties of cosmic rays from 
air shower observables.

The Observatory is located in the vicinity of the small city Malarg\"ue 
in Mendoza Province, Argentina, and began collecting data in 2004.  The 
construction of the fundamental design was completed by the end of 2008.  
Until October 2010 the Observatory has collected around 
$\unit[20\,000]{km^2\,sr\,yr}$ in exposure, which is significantly more 
than past cosmic-ray observatories combined. The Observatory is built 
around two types of detectors. Detectors on the ground sample air-shower 
particles as they arrive at the Earth's surface, while fluorescence 
detectors are measuring the light emitted when air-shower particles 
excite nitrogen molecules in the atmosphere.

The surface array \cite{operations} consists of 1600 fully autonomous 
surface detector (SD) stations, each being a light-tight tank filled 
with \unit[12]{t} of ultra-purified water observed by 3 photomultiplier 
tubes detecting the Cherenkov light produced as charged particles are 
traversing the water. The signals from the photomultipliers are read out 
with flash analog-to-digital converters with \unit[40]{MHz} sampling and 
stamped by the GPS time, allowing for detailed study of the arrival-time 
profile of shower particles. The tanks are placed on a triangular grid 
with a \unit[1.5]{km} spacing. The whole array covers an area of
$\unit[3000]{km^2}$. The surface array operates with close to a 100\% 
duty cycle, and the acceptance for events with energy above 
$\unit[3\!\times\!10^{18}]{eV}$ is nearly 100\% \cite{trigger}.

The fluorescence detectors (FD) \cite{fd} are placed in 4 buildings, 
each hosting 6 telescopes overlooking the surface array. Each telescope 
is equipped with $\unit[11]{m^2}$ segmented mirror, focusing the 
fluorescence light entering through a \unit[2.2]{m} diaphragm onto a 
camera made of 440 photomultiplier-tube pixels. The photomultiplier 
signals are sampled with \unit[10]{MHz}, delivering a time profile of 
the shower as it develops through the atmosphere. The FD can be operated 
only in darkness (night) with clear sky conditions (low aerosol and 
cloud coverage), and has a duty cycle of approximately 10 to 15\%. In 
contrast to the SD, the acceptance of FD depends strongly on the energy 
of the primary particle \cite{exposure}, and has an useful range 
extending down to around $\unit[10^{18}]{eV}$.

The two conceptually different detector systems provide complementary 
information about the particular air shower. The SD measures the lateral 
distribution and time structure of shower particles arriving at the 
ground, while the FD measures the longitudinal development of the shower 
through the atmosphere. Only a relatively small subset of showers is 
observed simultaneously by the SD and FD. These ``hybrid'' events are 
providing an invaluable calibration tool (see Fig.~\ref{f:spectrum} -- 
left). Particularly, the FD is performing a roughly colorimetric 
measurement of the shower energy since the amount of emitted 
fluorescence light is proportional to the deposited energy. On the other 
hand, the SD is extracting the shower energy through analysis of 
particle densities at the ground. These rely strongly on predictions of 
hadronic interaction models. Furthermore, we have to use model 
predictions describing physics at energies far beyond those accessible 
to current accelerator experiments, where the models have actually been 
tuned. Hybrid events therefore offer much more reliable estimate for a 
model-independent energy scale of the SD array. This is the crucial 
point in the successful design of the Pierre Auger Observatory since the 
SD has a much greater data sample than the FD due to the greater live 
time and coverage.

\section{Energy Spectrum}

\begin{figure}[t]
\centering
\includegraphics[width=\textwidth]{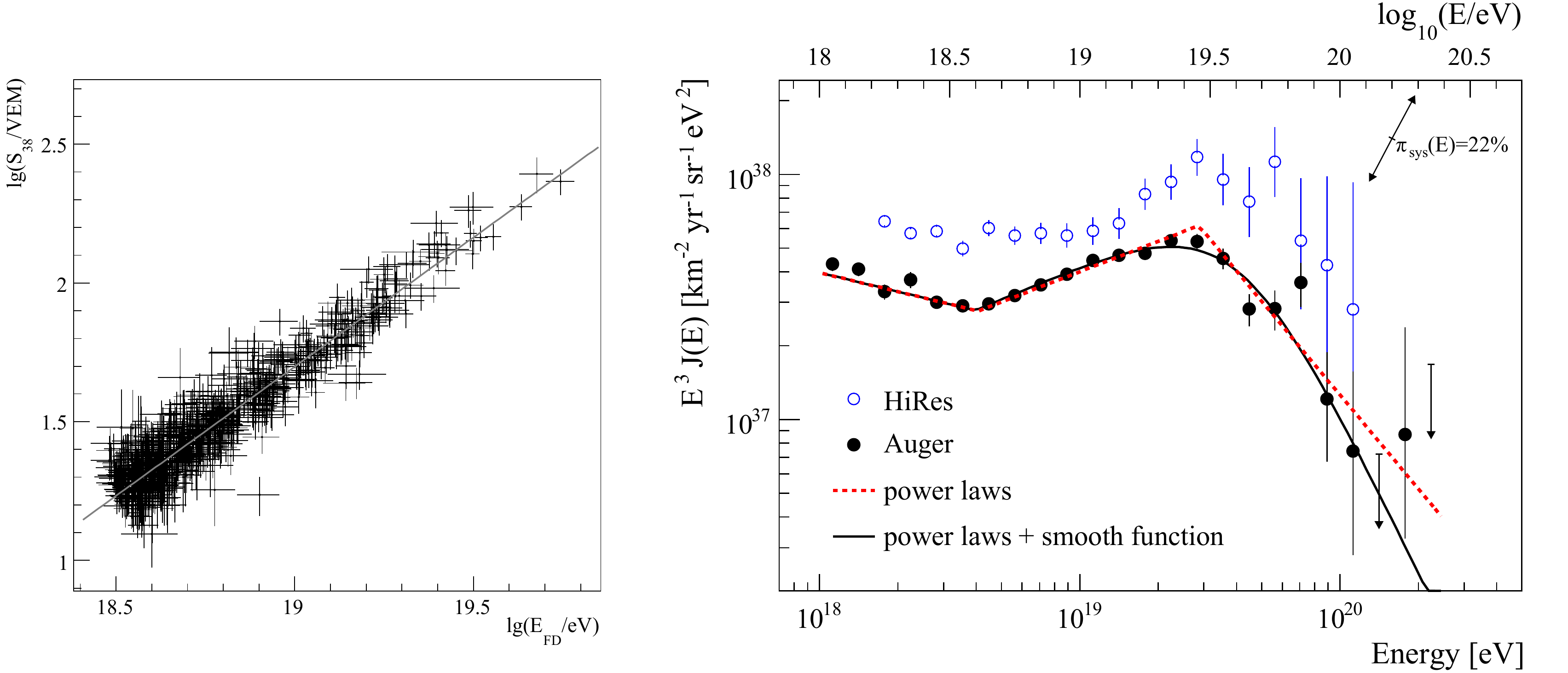}
\caption{\emph{Left:} The $S_{38^\circ}$ surface-detector energy 
estimator vs.\ the energy measured by the fluorescence detector for a 
sample of 795 high-quality hybrid events used to calibrate the 
surface-detector energy estimator. \emph{Right:} Combined energy 
spectrum from hybrid and surface-detector events \cite{spectrum}. The 
flux is multiplied by $E^3$ to straighten the otherwise steeply falling 
spectrum. The spectrum is compared to the HiRes results \cite{hires} 
(open circles). The results of the two experiments are consistent within 
the systematic uncertainties (two-sided arrow).}
\label{f:spectrum}
\end{figure}

Ultra-high energy cosmic-ray energy spectrum is one of the key parts in 
understanding the their origin and acceleration processes. These 
energies are up to eight orders of magnitude higher (or more than one 
order of magnitude higher in the center-of-mass energy) than those 
available from human-made accelerators like the LHC.

As noted above, the most reliable measurement of the primary energy is 
done by the observation of the fluorescence emission along the shower 
development path. Controlled atmosphere measurements in laboratories 
have determined the absolute yield of fluorescence photons per unit of 
energy deposit, their emission spectrum, and dependence on atmospheric 
parameters \cite{ave,monasor}. Once the shower geometry is reconstructed 
from the projected light trace and timing of the FD telescope pixels and 
SD stations, the absolute light intensity as a function of atmospheric 
depth $X$ can be calculated. The total energy deposited in the 
atmosphere is obtained by the integral of ${\rm d}E/{\rm d}X$ over depth 
$X$ (see details in \cite{unger}).

For SD events the shower arrival direction is reconstructed from the 
relative timing of the signals in SD stations and proceeds with a fit of 
the lateral distribution of particle densities at ground. An estimator 
$S(1000)$ denotes the signal size at \unit[1000]{m} from the shower core 
which was chosen to minimize the sensitivity to the shower-to-shower 
fluctuations and the unknown primary mass. $S(1000)$ will depend on the 
amount of atmosphere traversed by the shower and its attenuation is 
accounted for. The measured value $S(1000)$ is related to that expected 
at a chosen nominal zenith angle, in our case $38^\circ$. The new
quantity, $S_{38^\circ}$, measured in equivalent units of a signal from 
a vertical muon (VEM) is plotted in Fig.~\ref{f:spectrum} (left) vs.\ 
the absolute energy reconstructed from the FD. A fit of this correlation 
function then provides suitable energy calibration of the SD estimator.

The derived energy spectrum is presented in Fig.~\ref{f:spectrum} 
(right). Simple power law $E^{-\gamma}$ fits indicate that the ``ankle'' 
is located at $\log E=18.61\pm0.01$ and the spectral break at $\log 
E=19.46\pm0.03$ with the power-law indices being $\gamma=3.26\pm0.04$, 
$2.59\pm0.02$, and $4.3\pm0.2$ for the different regions, respectively.  
Within systematic uncertainties of about 20\% for the determination of 
the primary energy, the Auger and HiRes \cite{hires} spectrum can be 
considered consistent. The usual interpretation of a spectral 
suppression at this energy is the Greisen-Zatsepin-Kuz'min (GZK) effect 
\cite{gzk} and the loss of energy of heavy nuclei through the 
photo-disintegration on the cosmic microwave background (CMB).  
Nevertheless, the suppression could also be a result of cosmic-ray 
sources reaching their limits of acceleration. The ankle can be 
interpreted as the crossing point between the galactic and extragalactic 
origin of cosmic rays or, alternatively, as a feature caused by the pair 
production on CMB. This open question can be resolved by the 
measurements of the photon and neutrino fluxes predicted by the GZK 
effect.

\section{Mass composition}

\begin{figure}[t]
\centering
\includegraphics[width=\textwidth]{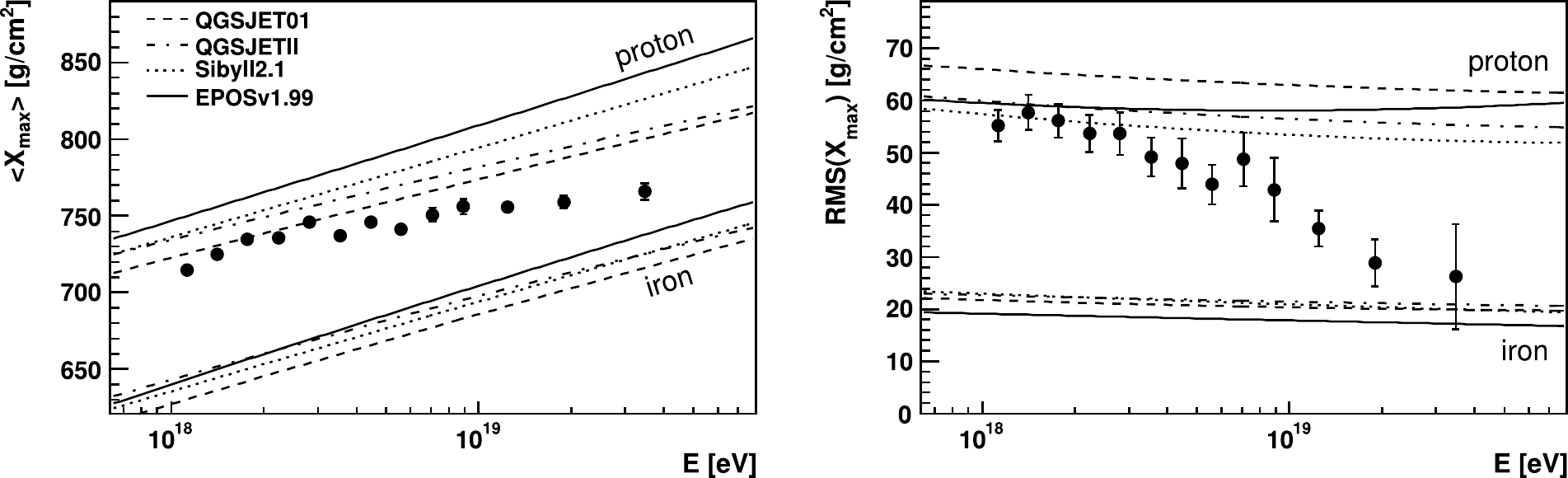}
\caption{Mean, $\langle X_\text{max}\rangle$, and the quadratic mean, 
$\operatorname{RMS}(X_\text{max})$, of the $X_\text{max}$ distribution 
as a function of energy \cite{xmax}. Data points are compared with 
expectations for proton and iron primary particles assuming four 
representative hadronic interaction models.}
\label{f:xmax}
\end{figure}

We have published results of FD measurements of the mean, $\langle 
X_\text{max}\rangle$, and the quadratic mean, 
$\operatorname{RMS}(X_\text{max})$, of the shower maximum $X_\text{max}$ 
distributions as a function of energy \cite{xmax}.

With FD detector the depth of shower development maximum, 
$X_\text{max}$, is directly viewed for many showers. Nevertheless, with 
such measurements collective quantities have to be formed carefully in 
order not to bias the sample with the event selection and/or 
reconstruction, resulting in deviations of $X_\text{max}$ distributions 
from the reality. One of the more important potential biases is the 
limited observation range in elevation (also known as field-of-view).  
Each telescope views the sky ranging in elevation from $\sim2^\circ$ to 
$\sim30^\circ$ above horizon.
%
%
Due to this limited field-of-view, showers occurring close
to the telescope are not selected if they have a shallow $X_\text{max}$
so that the $X_\text{max}$ distribution can exhibit bias towards deeper 
showers.
On the contrary, distant high-energy showers can be biased in the other 
direction, since deeply penetrating showers can hit the ground before 
the maximum is reached.

To study the magnitude and remove any potential biases the interior 
volume of the detecting medium is defined, excluding the external 
portion of the atmosphere, and applied to the data based on the shower 
geometry and energy. In these \emph{fiducial volume cuts} a minimum 
range of viewable atmospheric depths is imposed so that all possible 
$X_\text{max}$ values are certainly detected and measured with good 
accuracy \cite{unger2}. With these cuts, the analysis shows absence of 
bias using simulations of the detection and reconstruction of proton and 
iron primary particles (and their mixtures), so that $\langle 
X_\text{max}\rangle$ and $\operatorname{RMS}(X_\text{max})$ are well 
reproduced in the simulated shower sample. The typical measurement 
uncertainty of $X_\text{max}$ is obtained from the simulations and also
verified by the real events. For sufficiently high energies (above 
$\unit[10^{19}]{eV}$) a large fraction of showers are seen by two or 
more FD sites, so that $X_\text{max}$ can be measured independently.  
From this data we find that the typical measurement resolution around 
energy of $\unit[10^{19}]{eV}$ is $\unit[20\pm2]{g/cm^2}$, which is in 
excellent agreement with the simulation result of 
$\unit[19\pm0.1]{g/cm^2}$ \cite{xmax}. This gives ground to trusting 
simulated measurement resolutions used at lower energies. The 
$\operatorname{RMS}(X_\text{max})$ values given here are in quadrature 
subtracted by the detector resolution so that only the intrinsic 
variations are provided.

In Fig.~\ref{f:xmax} the behavior of the mean and RMS of the 
$X_\text{max}$ distribution is shown as a function of energy, compared 
with expectations from hadronic interaction models for proton and iron 
primary particles. The results favor a break in the elongation rate 
(defined as base-10 logarithmic slope $D_{10}={\rm d}X_\text{max}/{\rm 
d}\log E$) at an energy of $\unit[10^{18.25\pm0.05}]{eV}$, close to the 
position of the ``ankle'' in the energy spectrum given above. For higher 
energies, the elongation rate becomes smaller, $\unit[24\pm3]{g/cm^2}$ 
per decade, and is associated with a decreasing 
$\operatorname{RMS}(X_\text{max})$, both suggesting a possible change in 
the mass composition towards heavier nuclei.

Nevertheless, interpretation of these results requires comparison with 
simulations of air-shower development and the particular models for 
hadronic interactions employed by such simulations. In this light, the 
results described above can be (not very likely but still marginally 
probably) interpreted as composition being dominated by relatively light 
particles (protons) at highest energies but with some extreme changes in 
the underlying physics of interactions. All currently available models 
for these interactions are fits and extrapolations of accelerator data 
far beyond their nominal energies, e.g.\ the proton-air cross-section is 
derived from the $p$-$p$ cross-section at center-of-mass energies of up 
to several $\unit[10^{13}]{eV}$, equivalent to a fixed-target energy of 
several $\unit[10^{15}]{eV}$. Investigations of the influence of changes 
to standard extrapolations of cross-section, multiplicity, and 
elasticity on air shower observables have shown \cite{ulrich}
that it is easier to affect the mean of $X_\text{max}$ than to modify 
its fluctuations, and that out of the three parameters considered, the 
cross-section has the largest impact on 
$\operatorname{RMS}(X_\text{max})$, while extremely implausible changes 
are required to force $\unit[10^{19}]{eV}$ proton showers appear like 
iron showers under current simulations.

\begin{figure}[t]
\centering
\includegraphics[width=0.7\textwidth]{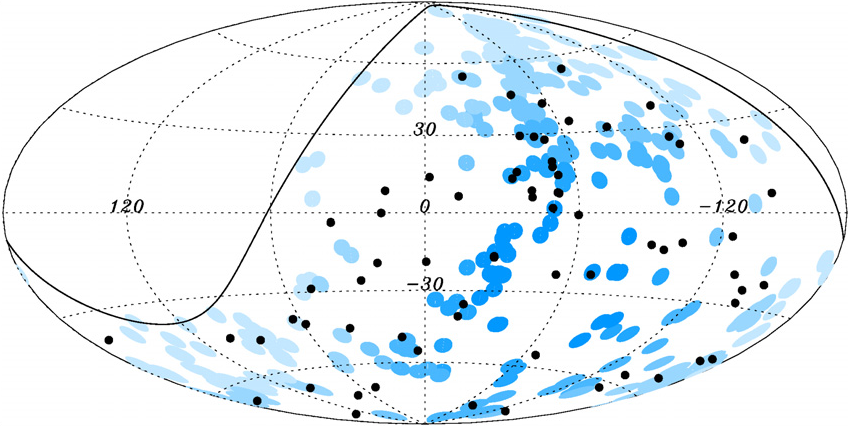}
\caption{The 69 arrival directions of cosmic rays with energy larger 
than $\unit[5.5\!\times\!10^{19}]{eV}$ detected until 31 December 2009 
are plotted as black dots in an Aitoff-Hammer projection of the sky in 
galactic coordinates. The solid line represents the border of the field 
of view of the Observatory for events with zenith angles smaller than 
$60^\circ$. Blue circles of radius $3.1^\circ$ are centered at the 
positions of the 318 active galactic nuclei in the V\'eron-Cetty-V\'eron 
catalog that lie within \unit[75]{Mpc} and that are within the field of 
view. Darker blue indicates larger relative exposure. The 
exposure-weighted fraction of the sky covered by the blue circles is 
21\%.}
\label{f:arrival_directions}
\end{figure}

\begin{figure}[t]
\centering
\includegraphics[width=0.75\textwidth]{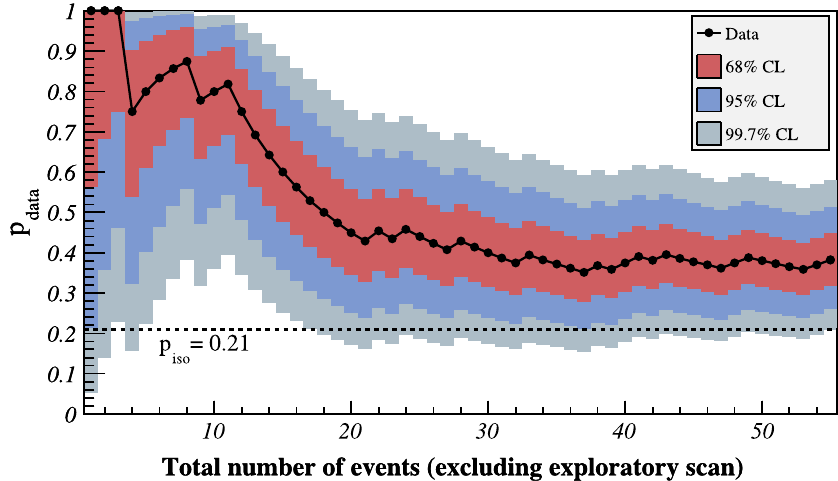}
\caption{The most likely value of the degree of correlation is plotted 
with black dots as a function of the total number of time-ordered events 
(excluding those in exploratory period).  The 68\%, 95\% and 99.7\% 
confidence level intervals around the most likely value are shaded. The 
horizontal dashed line shows the isotropic value $p_\text{iso}=0.21$.  
The current estimate of the signal is $0.38_{-0.06}^{+0.07}$.}
\label{f:degree_of_correlation}
\end{figure}

\begin{figure}[t]
\centering
\includegraphics[width=0.65\textwidth]{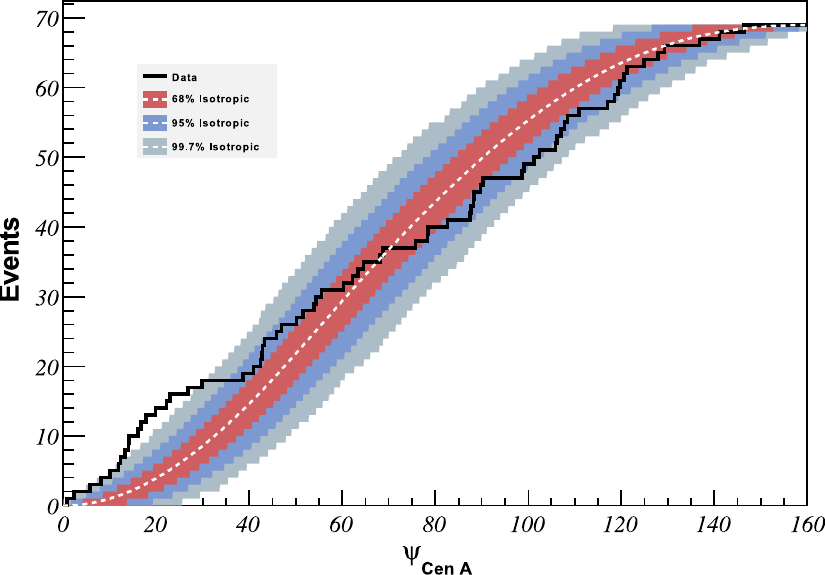}
\caption{Cumulative number of events with energy larger than
$\unit[5.5\!\times\!10^{19}]{eV}$ as a function of angular distance from 
the direction of Centaurus A galaxy (also known as NGC 5128). The bands 
correspond to the 68\%, 95\% and 99.7\% dispersion expected for an 
isotropic flux.}
\label{f:cen_a_correlation}
\end{figure}

\section{Arrival Directions}

Assuming that the highest-energy cosmic rays observed by the Pierre 
Auger Observatory are according to our current knowledge on the galactic 
and intergalactic magnetic fields exhibiting relatively unperturbed 
trajectories (a modus also known as the \emph{cosmic-ray astronomy}), it 
is interesting to check if any anisotropy begins to emerge at these high 
energies so that the potentially uneven distribution of sources can be 
revealed. Furthermore, if the observed flux suppression is really a 
consequence of the GZK effect, there is an associated GZK horizon of the 
order of \unit[100]{Mpc}, beyond which cosmic rays with starting 
energies near $\unit[10^{20}]{eV}$ will be observed with much smaller 
energies. Since the matter density within our local cosmological 
neighborhood of about \unit[100]{Mpc} is not isotropic, this opens a 
possibility to potentially detect the anisotropy in the recorded data 
sample. We have performed point source studies as well as harmonic 
analysis of arrival directions, which both characterize anisotropy at 
various angular scales~\cite{harmonic}.

One way to increase the chance of finding potential sources of 
ultra-high energy cosmic rays is to check for correlations between 
cosmic-ray arrival directions and known positions of interesting 
astrophysical candidates. However, care must be taken to appropriately 
take into account the reduction of statistical significance due to the 
repeated trials made in such procedures. Due to this, the Pierre Auger 
Collaboration decided to follow a predefined process. As first, an 
exploratory scan of the correlation between the data and various source 
catalogs has been performed, optimizing various parameters and cut 
choices. The results of this exploratory period were then used to design 
prescriptions used for the subsequently gathered data.

The resulting prescription was designed to test the correlation of
events with energies larger than $\unit[5.6\!\times\!10^{19}]{eV}$ with 
objects in the V\'eron-Cetty \& V\'eron catalog of active galactic 
nuclei. The prescription implied a search of $3.1^\circ$ large windows 
around nearby catalog objects with redshifts $z<0.0018$. The 
significance threshold set in the prescription was met in 2007 
\cite{agn,agn2} with 9 out of the 13 events in the sample correlating.
The number of correlating events is now 21 out of 55, i.e.\ the 
correlating fraction is $0.38^{+0.07}_{-0.06}$ with 0.21 expected for 
isotropically distributed events \cite{update}. A sky-map showing the 
locations of the events with energies above the cut is displayed in 
Fig.~\ref{f:arrival_directions}. The fraction of correlating events for 
the prescription period and for the following period is shown in 
Fig.~\ref{f:degree_of_correlation}. Compared to the initial results, the 
strength of the correlation appears to fall after the first 20 or so 
events, however, evidence for anisotropy currently remains at a stable 
fraction of $\sim0.35$ with only $0.3\%$ to find \emph{by chance} 21 or 
more of 55 events from an isotropic distribution correlating under these 
parametric conditions.

We have observed a number of other interesting correlations 
\cite{update}, including comparisons with other catalogs of 
astrophysical objects as well as a specific search around the direction 
towards Centaurus A (see Fig.~\ref{f:cen_a_correlation}). The maximum
departure from isotropy occurs for a ring of $18^\circ$ around the
object, in which 13 events are observed compared to an expectation of 
only 3.2 from isotropy. While these events could be coming from 
Centaurus A itself, which is only about \unit[4]{Mpc} away, it is
also possible that they originate in the Centaurus galaxy cluster at a 
distance of about \unit[45]{Mpc}. Nevertheless, it is important to keep 
in mind that these are all \emph{a posteriori} studies, an therefore 
without reliable confidence levels for anisotropy since the number of 
trials is unknown.

\acknowledgments
Author would like to acknowledge the support from the Slovenian Ministry 
for Higher Education, Science, and Technology, and Slovenian Research 
Agency, and thank all the colleagues from the Pierre Auger collaboration 
for being involved and working together on this magnificent project.

\end{document}